\documentclass[11pt,twoside]{article}
\usepackage{asp2004}
\usepackage{psfig}
\usepackage{epsf}
\usepackage{graphics}
\usepackage{lscape}
\def\edcomment#1{\iffalse\marginpar{\raggedright\sl#1\/}\else\relax\fi}
\reversemarginpar

\def\kms{\ifmmode {\,\rm km \, s^{-1}}
\else {$\rm km \, s^{-1}$}\fi}
\def\Mpc{\ifmmode {\, h^{-1} \, {\rm Mpc}}
\else {$h^{-1}\,$ Mpc}\fi}

\def\s8{{\sigma_8}}

\def\ltsima{$\; \buildrel < \over \sim \;$}
\def\simlt{\lower.5ex\hbox{\ltsima}} 
\def\gtsima{$\; \buildrel > \over \sim \;$} 
\def\simgt{\lower.5ex\hbox{\gtsima}}

\def\omegam{{\Omega_{\rm m}}}

\def\omegab{{\Omega_{\rm b}}}

\def\omegal{{\Omega_\Lambda}}
 
\def\omegabh2{{\omegab h^2}}

\def\s8m{{\sigma_{8{\rm m}}}}
\def\s8g{{\sigma_{8{\rm g}}}}

\def\hompc{\ifmmode {\,h\,\rm Mpc^{-1}}
\else {$h^{-1}$~Mpc}\fi}
\def\m@th{\mathsurround=0pt }
\def\eqalign#1{\null\,\vcenter{\openup1\jot \m@th
 \ialign{\strut\hfil$\displaystyle{##}$&$\displaystyle{{}##}$\hfil
 \crcr#1\crcr}}\,}

\begin{document}
\title{Quantifying the Cosmic Web in the New Era of Redshift Surveys} 
\author{Ofer Lahav}
\affil{Department of Physics and Astronomy , University College London, 
Gower Street, London WC1E 6BT, UK\\}

\begin{abstract}
Two main strategies have been implemented in mapping the local universe:
whole-sky `shallow' surveys  and `deep' surveys
over  limited parts of the sky.
The two approaches complement each other
in studying cosmography and statistical properties of the Universe. 
We summarise  some results on the power spectrum of fluctuations 
and Wiener reconstruction of the density field from 
the 2dF Galaxy Redshift Survey (2dFGRS) of 230,000 redshifts.
We then discuss future challenges in quantifying the web of cosmic structure
in the on-going redshift surveys.
\end{abstract}
\thispagestyle{plain}

\section{Introduction}

The recent measurements of the Cosmic Microwave Background
fluctuations, supernovae Ia, redshift surveys, clusters of galaxies
and other probes suggest a `concordance' model in which the  universe
is flat and contains
approximately 4\% baryons, 26\% cold dark matter and 70\% dark
energy. It remains to be seen if this model will survive future tests,
but in any case there are many challenges ahead to understand the
formation of galaxies, and how they trace the mass distribution in the
non-linear regime.  Redshift surveys provide  an important
bridge between `linear cosmology' and the more complex processes of
galaxy formation.

Two main strategies have been implemented in mapping the local universe:
whole-sky `shallow' surveys (e.g. IRAS)  and 'deep' surveys
over a limited parts of the sky (e.g. 2dFGRS, SDSS).
The Table below summarises  the properties of the
main new surveys:
2dFGRS\footnote{http://www.mso.anu.edu.au/2dFGRS/},
SDSS\footnote{http://www.sdss.org/} + LRG\footnote
{Another part of the SDSS is the `Luminous Red Galaxies' (LRG)
with median redshift ${\bar z} \sim 0.5$,
An extension of the survey to higher redshift is now underway utilising
2dF.},
2MASS\footnote{http://www.ipac.caltech.edu/2mass/}
/6dFGS\footnote{http://www.mso.anu.edu.au/6dFGS/},
DEEP2\footnote{http://deep.berkeley.edu/}, and
VIRMOS\footnote{http://www.astrsp-mrs.fr/virmos/}.


\begin{center}
\begin{tabular}{|l||c|c|c|}
\hline
Survey
&number of galaxies
&median redshift
&angular coverage (sq. deg)
\\
\hline
2dFGRS & 230k & 0.1 & $\sim$ 1,800\\
SDSS & 1000k & 0.1 & $\sim$ 10,000\\
2MASS-2MRS & 25k & 0.02 & $\sim$ 40,000\\
2MASS-6dFGS & 150k & 0.05 & $\sim$ 20,000\\
DEEP2 & 65k & $\sim 1$ & 3.5 \\
VIRMOS &150k & $\sim 1$ & 16 \\
\hline
\end{tabular}
\end{center}


Each strategy has its pros and cons.
The whole-sky surveys have given useful 'full picture' of
the local cosmography 
and they have allowed us to predict the local velocity field assuming 
that light roughly traces mass.
The complete picture depends on careful mapping of the 
Zone of Avoidance (ZoA),
as discussed in detail at the Proceedings of this Cape Town (2004) 
meeting and at the previous two 
ZoA conferences in Paris (1994) and Mexico (2000).
The deep limited-sky surveys are very useful for statistical studies
such as the power spectrum.
Both types of surveys pose challenges for quantifying the web
of cosmic structure.

Using simulations Bond, Kofman \& Pogosyan (1996) coined the
term `cosmic web' and argued that a filament-dominated structure was
already present in the overdensity fields of the initial Gaussian
fluctuations, and was then amplified over a Hubble time by non-linear
gravitational dynamics.  In the new era of large redshift surveys
(see the Table) and huge simulations
the next important step is to quantify this `cosmic web' using various
novel statistical measures beyond the traditional methods
(e.g. Martinez \& Saar 2002; Lahav \& Suto 2004 for reviews)
and to identify `Great
Attractors', `Great Walls', `Zeldovich pancakes' and voids.  This will
allow us to understand the role of initial conditions vs.
non-linear gravitational evolution and to constrain cosmological
models and scenarios for biased galaxy formation.

\section{Results from the 2dF Galaxy Redshift Survey}

Redshifts surveys in the 1980s and the 1990s (e.g the CfA, IRAS and Las
campanas surveys) measured redshifts
of thousands to tens of thousands of galaxies.
Multifibre technology now allows us to measure redshifts of
millions of galaxies. 
The Anglo-Australian 2
degree Field Galaxy Redshift Survey\footnote{The 2dFGRS Team comprises:
      I.J. Baldry, C.M. Baugh, J. Bland-Hawthorn, T.J. Bridges, R.D. Cannon, 
      S. Cole,
      C.A. Collins,  
      M. Colless,
      W.J. Couch, N.G.J. Cross, G.B. Dalton, R. DePropris, S.P. Driver,
      G. Efstathiou, R.S. Ellis, C.S. Frenk, K. Glazebrook, E. Hawkins, 
      C.A. Jackson,
      O. Lahav, I.J. Lewis, S.L. Lumsden, S. Maddox, 
      D.S. Madgwick, S. Moody, P. Norberg, J.A. Peacock, B.A. Peterson,
      W. Sutherland, K. Taylor. 
For more details on the survey and resulting publications see http://www.mso.anu.edu.au/2dFGRS/}
(2dFGRS)  
measured redshifts for 230,000 galaxies
selected from the APM catalogue. The survey is now complete and publically available.
The median redshift of the
2dFGRS is ${\bar z} \sim 0.1$,  
down to an
extinction corrected magnitude limit of $b_J<19.45$ (Colless et al. 2001). 
A sample of this size allows large-scale structure statistics
to be measured with very small random errors. 
Here we summarize some recent results 
from the 2dFGRS on clustering and galaxy biasing.
Comprehensive recent reviews are given by Colless (2003) and Peacock (2003).

\subsection{The power spectrum of 2dF galaxies}

  An initial estimate of the convolved, redshift-space power spectrum of the
 2dFGRS has  been determined (Percival et al. 2001)
 for a sample of 160,000 redshifts. 
 On scales $0.02<k<0.15 \hompc$, 
 where $H_0 = 100 h$ km/sec/Mpc,
 the data are
 robust and the shape of the power spectrum is not affected by
 redshift-space or non-linear effects, though the amplitude
 is increased by redshift-space distortions.
 Percival et al. (2001), Efstathiou
 et al. (2002) and Lahav et al. (2002)  compared the  
 2dFGRS and CMB
 power spectra, and concluded that they are consistent with each other.

A key assumption in deriving cosmological parameters from redshift surveys is that 
the biasing parameter, 
defined as the ratio of 
 of galaxy to matter power spectra, 
is constant, i.e. scale independent.
On  scales of
$0.02 < k < 0.15 \hompc$ 
the fluctuations are close
to the linear regime, and there are theoretical  reasons 
(e.g. Fry 1996; Benson et al. 2000)
to expect that on large scales 
the biasing parameter 
should tend to a constant and close to unity at the present epoch. 
This is supported by the derived biasing close to unity by combining 
2dFGRS with the CMB (Lahav et al. 2002) and by the 
study of the bi-spectrum of the 2dFGRS alone (Verde et al. 2002).

The 2dFGRS power spectrum (Figure 1) was fitted in Percival et al. (2001) 
over the above range in $k$, 
assuming scale-invariant primordial 
fluctuations and a $\Lambda$-CDM cosmology, for 
four free parameters: $\omegam h$, $\omegab/\omegam$, $h$  
and the redshift space $\sigma^S_{8{\rm g}}$.
The amplitudes
of the linear-theory rms fluctuations are traditionally labeled  $\sigma_{8{\rm m}}$ 
in mass  $\sigma_{8{\rm g}}$ in galaxies, defined on $8 h^{-1}$ Mpc spheres.
Assuming a Gaussian prior on the Hubble constant $h=0.7\pm0.07$ (based
on Freedman et al. 2001) the shape of the recovered spectrum within
the above $k$-range was used to yield 68 per cent confidence limits on
the shape parameter $\omegam h=0.20 \pm 0.03$, and the baryon fraction
$\omegab/\omegam=0.15 \pm 0.07$, in accordance with the popular
`concordance' model (e.g. Bahcall et al. 1999; Lahav \& Liddle 2004).  
For fixed `concordance model' parameters $n=1, \omegam = 1 - \omegal = 0.3$,
$\Omega_{\rm b} h^2 = 0.02$ and a Hubble constant $h=0.70$, 
the amplitude of 2dFGRS galaxies in redshift space is $\sigma_{8{\rm
g}}^S (L_s,z_s) \approx 0.94$ (at the survey's effective luminosity and redshift).

Recently the SDSS team presented their results for the power spectrum
(Tegmark et al. 2003a,b; Pope et al. 2004), and they found good agreement
with the 2dFGRS gross shape of the power spectrum.
Pope et al. (2004) emphasize that SDSS alone cannot 
break the degeneracy between $\omegam h$ and $\omegab/\omegam$
because the baryon oscillations are not resolved given
the window function of the survey.

\subsection {Upper limits on the neutrino mass} 

Solar, atmospheric, and reactor neutrino experiments have confirmed
neutrino oscillations, implying that neutrinos have non-zero mass, but
without pinning down their absolute masses.  While it is established
that the effect of neutrinos on the evolution of cosmic structure is
small, the upper limits derived from large-scale structure could help
significantly to constrain the absolute scale of the neutrino masses.
Elgar\o y et al. (2002) used the 2dFGRS power spectrum (Figure 1) to
provide an upper limit $m_{\nu,\rm tot} < 2.2\;{\rm eV}$ ,
i.e. approximately 0.7 eV for each of the three neutrino flavours, or
phrased in terms of their contribution to the matter density,
$\Omega_{\nu} / \Omega_{\rm m} < 0.16$.

The WMAP team (Spergel et al. 2003)  reported an improved
limit of $m_{\nu,\rm tot} < 0.71\;{\rm eV}$ (95\% CL).  
Actually the main neutrino signature comes from
the 2dFGRS and the Lyman $\alpha$ forest which were combined with the
WMAP data. The main contribution of WMAP is that it constrains better the
other parameters involved, e.g. $\omegam$ (see also Hannestad 2003 and
Tegmark et al. 2003b for similar results from SDSS+WMAP).
Despite the uncertainties involved, it is remarkable that the results from
redshift surveys give upper limits which are lower than 
those deduced from laboratory experiments, e.g. tritium decay.

\begin{figure}
\plotone{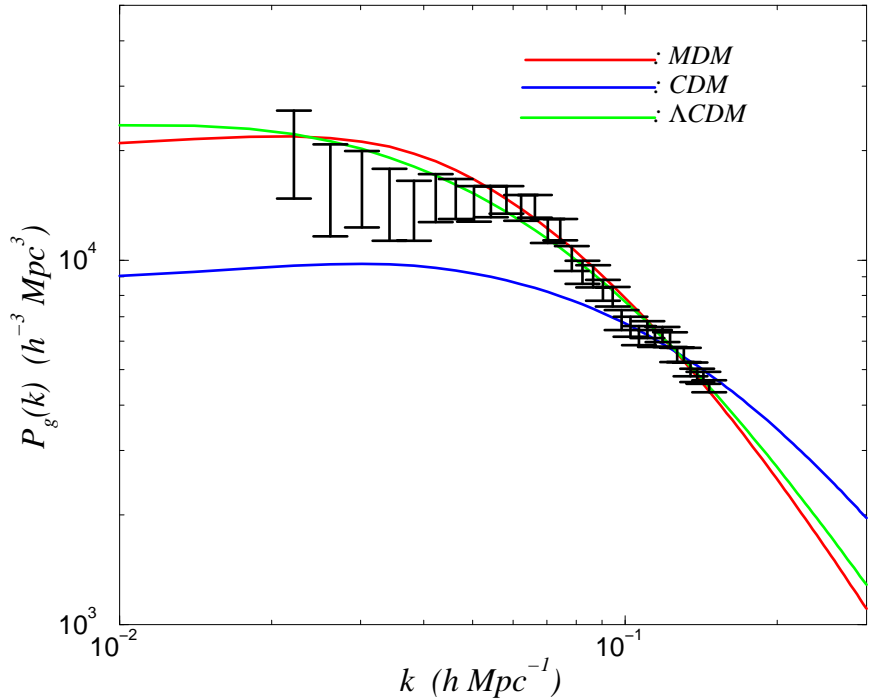}
\label{pk_mdm}
\caption{The observed 
2dFGRS power spectrum (in redshift space and convolved  with the 
survey window function; Percival et al. 2001) contrasted 
with models.
The three models are 
the old Cold Dark Matter  model
($\Omega_{\rm m}=1$, $\Omega_\nu=0$, $h=0.45$, $n=0.95$)
which poorly fits the data,
the `concordance' model  
$\Lambda {\rm CDM}$ ($\Omega_{\rm m}=1$, $\Omega_\Lambda=0.7$, 
$\Omega_\nu = 0$, $h=0.7$, $n=1.0$) 
and  Mixed Dark Matter
 ($\Omega_{\rm m}=1$, $\Omega_\nu=0.2$, $h=0.45$, $n=0.95$),
all with $\Omega_{\rm b}h^2 = 0.024$.
The models were normalized  to each 
data set separately, but otherwise these are assumed models, not 
formal best fits.
Only the range $ 0.02 < k < 0.15 \hompc$  is used 
at the present linear theory analysis.
These scales of $k$ 
roughly correspond to CMB harmonics $200 <  \ell < 1500$
in a flat $\omegam = 0.3$ universe.
From Elgar\o y \& Lahav (2003).}
\end{figure}

\section{Wiener reconstruction of 2dFGRS}

Wiener filtering is a well-known technique and it has been applied to
many fields in astronomy. For example, the method was used to
reconstruct the angular distribution over the whole sky including the
ZoA (Lahav et al. 1994), the real-space density, velocity and
gravitational potential fields of the 1.2-Jy IRAS (Fisher et
al. 1995).  The Wiener filter was also applied to the reconstruction
of the  maps of the cosmic microwave background temperature
fluctuations.  A detailed formalism of the Wiener filtering method as
it pertains to the large-scale structure reconstruction can be found
in Zaroubi et al. (1995).  The Wiener filter is optimal in the sense
that the variance between the derived reconstruction and the
underlying true density field is minimised. As opposed to ad hoc
smoothing schemes, the Wiener filtering is
determined by the data. In the limit of high signal-to-noise, the
Wiener filter modifies the observed data only weakly, whereas it
suppresses the contribution of the data contaminated by shot noise.

Erdogdu et al. (2004)  reconstructed 
the underlying density field of the Two-degree Field
Galaxy Redshift Survey (2dFGRS) for the redshift range 
$0.04 < z < 0.20$
using the Wiener filtering method. 
They used a
variable smoothing technique with two different effective resolutions:
5 and 10 $h^{-1}$ Mpc at the median redshift of the survey. 
They identified all 
major superclusters and voids in 2dFGRS. In particular, they found
two large superclusters and two large local voids. 
One of the two large superclusters is shown in Figure 2.
The full set of
colour maps can be viewed on the World Wide Web at
http://www.ast.cam.ac.uk/$\sim$pirin.
For comparison see the catalogue of superclusters derived by Einasto et al. 
(2003) from the SDSS.

\begin{figure}
\plotone{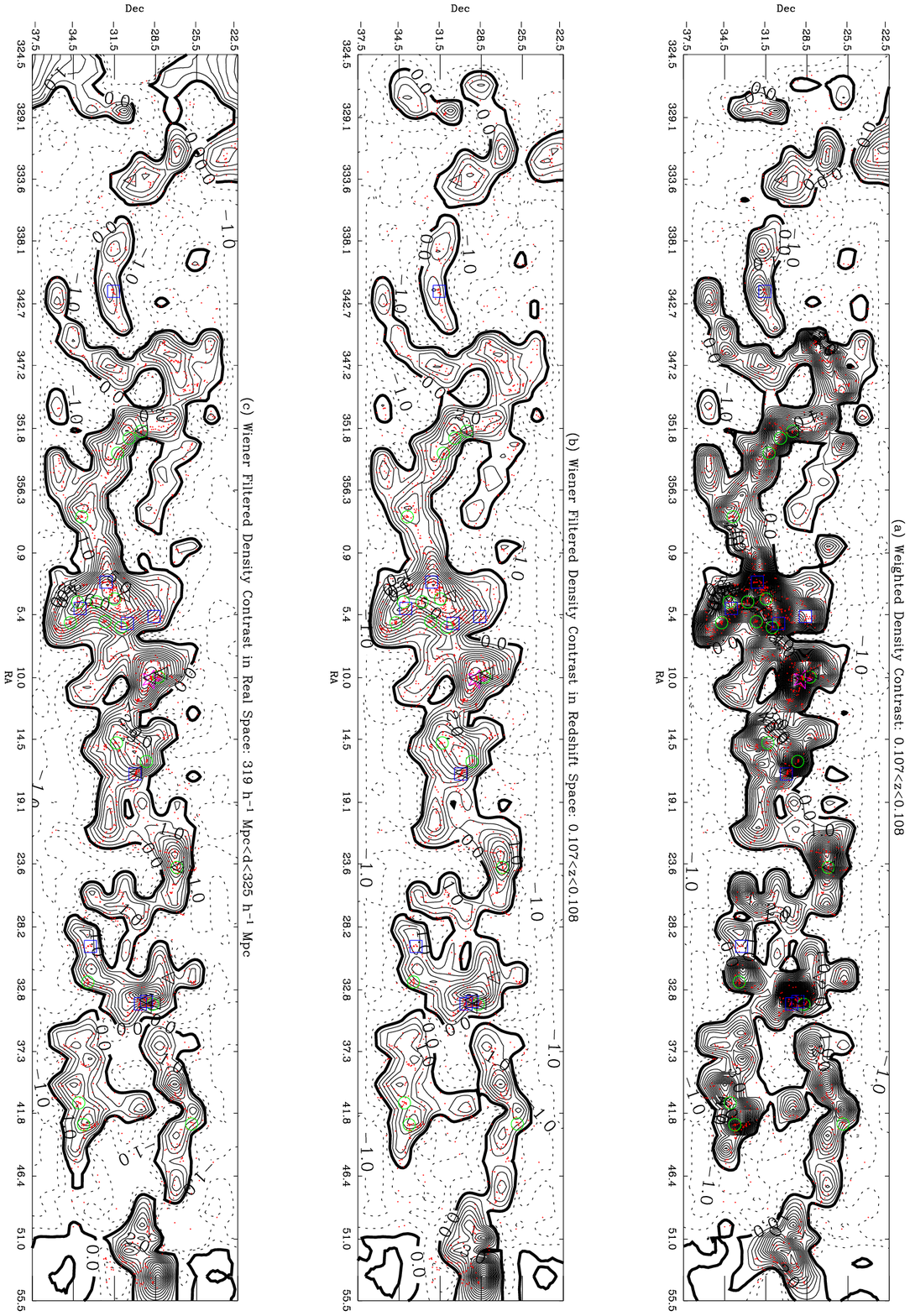}
\label{pirin1}
\caption{Wiener reconstruction of a 2dFGRS  redshift shell
$0.107 < z < 0.108$
for $5 h^{-1}$ Mpc cells.
(a) the redshift-space density field weighted by the selection function and the
 angular mask; (b) the Wiener filtered density field 
in redshift space; (c)  the Wiener filtered density field 
in real space, i.e. after correction for redshift distortion. 
The contours are spaced at $\Delta \delta=0.5$
with solid/dashed lines denoting positive/negative contours;
the heavy solid lines correspond to $\delta=0$.
The connectivity of structure across the field is striking.
The Supercluster centred at $RA \approx 2^\circ; Dec \approx -31^\circ$
is one of the two largest superclusters in the 2dFGRS, and it contains
20 Abell clusters and approximately 80
smaller groups.
From Erdogdu et al. (2004).}
\end{figure}

\section{Discussion: Future studies of the cosmic web}

We motivate the great need for new approaches to
analysis of redshift surveys and simulations by some illustrative examples:

\noindent
(i) Consider two images, say of a cat and a dog.
If we take Fourier transforms of both, and swap the
amplitudes and phases, we will still be able to recognise the
cat in the image which retains its original phases, even if it has
the amplitudes from the dog's original image!
Phase information is thrown away  in
the commonly used power spectrum (or its Fourier
transform, the two-point correlation function).
Two realisations of the galaxy distribution may have the same
power spectrum, but they may look very different due to phase correlations.
These phase correlations are expected to be due to the non-linear effects
in the evolution of the gravitational instability
and would arise even
if the primordial fluctuations were purely Gaussian.

\noindent
(ii) The 2-degree-Field Galaxy Redshift Survey (2dFGRS) power spectrum is consistent with the
low density ($\Omega_{\rm m} = 0.3$) Cold Dark Matter model (Percival et al. 2001).
Volume averaged $p$-point correlation functions
 up to order $p=6$ have recently been calculated
(Baugh et al. 2004)
and in particular on small scales they are sensitive to the appearance
of two rich superclusters in the 2dFGRS volume, one of them is shown
in Figure 2.  Visual inspection of the Abell
catalogue suggests that such superclusters are quite common. However,
they seem less common in $\Lambda$-CDM simulations which nevertheless
do agree with the 2dFGRS power spectrum ($p=2$) statistic.  It is
important to know whether  the $\Lambda$-CDM simulations pass the test of
high order moments, whether they require strong biasing in high density
regions and whether 2dFGRS is a fair sample of the nearby universe.

\noindent
(iii) Recent studies of higher moments, e.g. the three point
correlation function (or its Fourier transform, the bi-spectrum)
in both 2dFGRS and SDSS
showed that they set important constraints on biasing, although the
interpretation is model dependent (e.g. Verde et al. 2002; Kayo et
al. 2004).
\noindent

\bigskip

It is  timely to address these issues now for a number of reasons:



\noindent
$\bullet$
 For the first time the surveys are large enough to extract
 volume limited subsets. Some of the statistics (e.g. minimal
 spanning tree, percolation and Minkowski functionals) 
 could not be applied effectively
 to flux limited surveys, where the mean separation between
 observed  objects varies with distance from the observer.
 With volume limited samples we can apply these and new methods
 easily, and contrast data with simulations on equal footing.

\noindent
 $\bullet$
The new surveys are also large enough now to sub-divide the galaxies
by colour or spectral type. Recent studies of 2dFGRS and SDSS
show the bimodality of galaxy populations in colour or related properties
(e.g. Madgwick et al. 2002;
Kauffmann et al. 2004)
and that clustering patterns of `red' and `blue' galaxies
are quite different on scales smaller than 10 $h^{-1}$Mpc
(e.g. Zehavi et al. 2003;  Madgwick et al. 2003,
Wild et al. 2004).

\noindent
$\bullet$
We can now probe the evolution of clustering patterns with redshift,
given DEEP2 and VIRMOS at redshift ${\bar z} \sim 1$ compared
to  2dFGRS and SDSS
at $ {\bar z} \sim 0.1$.

\noindent
$\bullet $
The N-body simulations (e.g. Virgo)
are now well advanced in resolution and volume to allow
us to produce 2dF-like or SDSS-like samples.


\section{Acknowledgements}

I thank 
the conference organisers for the hospitality in Capetown,  
and  \O ystein Elgar\o y and Pirin Erdogdu
for their contribution to the work presented here.
I acknowledge a PPARC Senior Research Fellowship.

 \label{lastpage}

\end{document}